\documentclass[preprintnumbers,nofootinbib,noshowpacs,eqsecnum,twocolumn,prd]{revtex4}

\usepackage{graphicx}
\usepackage{amsmath,amsthm,amssymb}
\usepackage{mathrsfs,bbm}
\usepackage{array,color}
\usepackage{tikz}

\makeatletter       

\renewcommand{\p@subsection}{}

\renewcommand{\eqnum}{\refstepcounter{equation}\textup{\tagform@{\theequation}}} 
\makeatother

\newcommand*{\unitmatrix}{\mathbbm{1}}

\newcommand*{\tvec}[1]{\boldsymbol{#1}}              
\newcommand*{\trans}{\mathrm{T}}                     

\DeclareMathOperator{\diag}{diag}		

\begin{document} 

\author{Markos Maniatis}
\affiliation{Centro de Ciencias Exactas \& Departamento de Ciencias B\'asicas,  Universidad del B\'io-B\'io,  Chill\'{a}n,  Chile}
\email{maniatis8@gmail.com}

\title{The tree world}

\begin{abstract}Tree amplitudes of any gauge theory and gravity can
be factorized into primitive three-particle amplitudes by the 
BCFW recursion relations. We show that the amplitudes at
any perturbation order are given by tree amplitudes with additional pairs
of particles in the forward limit. As an example we compute the four-gluon all-plus helicity amplitude at next-to-leading order.
\end{abstract}

\maketitle
\flushbottom

\section{Introduction}
\label{sec:intro}

It is well known that the computation of scattering amplitudes via Feynman diagrams 
has a number  of drawbacks: First of all, Feynman diagrams are plagued with gauge redundancies, that is, single diagrams are in general not gauge invariant and only the sum of diagrams (or certain subsets) restore gauge invariance. 
In particular, the three-gluon vertex violates gauge invariance in general and an additional four-gluon vertex has to be introduced along with ‘‘ghost'' particles. Feynman diagrams are in general even not Lorentz covariant, and only the product with polarization ‘‘tensors'' form Lorentz-invariant expressions. 
Moreover, in the diagrams off-shell particles have to be introduced, that is, virtual particles. 

Of course, Feynman diagrams are nothing more than a mathematical prescription 
to generate amplitudes. However, this is also done in a redundant way: 
Every Feynman diagram to a certain perturbation order is of the same order as the final result.
This means that most contributions cancel each other. Maybe the most striking 
example in this respect is the Parke-Taylor formula for the tree level scattering amplitude of $n$ external gluons, two carrying negative helicity and the rest positive helicities (maximally helicity violating amplitude). Suppressing the color factor and the coupling constant, the correct result was conjectured~\cite{Parke:1986gb},
\begin{equation} \label{Parke-Taylor}
A_n(1^+, \ldots, i^-, \ldots, j^-, \ldots, n^+)
=
\frac{ \langle i j \rangle}
{\langle 12 \rangle \langle 23 \rangle \cdots \langle n1 \rangle}\;,
\end{equation}
where the expressions $\langle ab \rangle$ denote scalar products in the helicity formalism - see appendix~\ref{app:spinor} for a very brief recap and the conventions used.
In terms of Feynman diagrams the computation of this amplitude requires for $n=5$ external gluons to calculate 25 diagrams, and for $n=6$ external gluons already 220 diagrams.
 
 A milestone in the program to deduce scattering amplitudes from 
{\em first principles}, that is, Lorentz invariance, locality, unitarity, and gauge groups, 
are the BCFW recursion relations~\cite{Britto:2004ap,Britto:2005fq}.
Via analytic continuation, or deformation, of the momenta of external particles it has been 
shown that in gauge theories and gravity the tree amplitudes can be factorized into 
on-shell {\em primitive} three-particle amplitudes. The BCFW recursion relations turn out to be very powerful. For instance the Parke-Taylor formula~\eqref{Parke-Taylor} can be proven by the BCFW recursion relations.

However, the BCFW recursion relations rely on tree amplitudes, which of course form only an unphysical subset of the scattering amplitude. There has been spent a lot of effort to go beyond leading order. 
In particular at next-to-leading order it has been shown that via unitary cuts the amplitudes can be computed~\cite{vanNeerven:1983vr, Bern:1994zx, Bidder:2005ri, Anastasiou:2006jv, Giele:2008ve}.

A great step towards the derivation of amplitudes at any perturbation order is the 
Feynman-tree theorem~\cite{Feynman:1963ax,Feynman::FTT}.
In a Feynman diagram at any perturbation order a loop can be {\em opened} by cutting the propagators, which form the loop, in all different ways. Recursively, all
loops can be opened and finally there remain only tree diagrams. This approach has
been modified and applied to different calculations of amplitudes~\cite{Catani:2008xa,Bierenbaum:2010cy,Baadsgaard:2015twa} and combined with a subsequent application of the BCFW recursion relations~\cite{Maniatis:2015kex,Maniatis:2016nmc,Maniatis:2019pig}.
However, this approach is dedicated to Feynman diagrams based on loops, that is, virtual particles. As to be expected, there is no direct translation from Feynman diagrams opening the loops and on-shell amplitudes at higher perturbation orders~\cite{Caron-Huot:2010fvq}.  

Here we want to propose an approach relying on on-shell amplitudes:
We argue that in the tree on-shell amplitudes we inevitably have to take 
pairs of particles in the forward limit into account. With the particles in the forward limit with opposite helicities and momenta, the amplitudes are physically indistinguishable from the leading oder on-shell tree amplitudes. We study on-shell tree diagrams adding systematically pairs of particles in the forward limit. In a second step we apply the BCFW recursion relations to these tree amplitudes 
taking particles in the forward limit into account. 
We show that we can factorize the on-shell amplitudes at any perturbation order into
primitive three-point on-shell amplitudes. This is in particular very attractive since
the three-point primitive amplitudes are fixed by Lorentz-invariance, that is, little group scaling and the gauge group.

We apply the proposed method to the four-gluon all-plus helicity amplitude at next-to-leading order. The four-point amplitude is an excellent framework since it is rather simple but illustrates the proposed method in detail. Eventually we sketch how at next-to-next-to leading order the four-gluon amplitude can be computed in the proposed formalism, taking particles in the forward limit into account.

\section{Tree amplitudes}
\label{sec:tree}

We start considering an arbitrary $n$-point on-shell tree amplitude,
\begin{equation}
A_n(1^{h_1},2^{h_2}, \ldots, n^{h_n}).
\end{equation}
We note that we are considering here the complete amplitude and not a color-ordered subamplitude. The momenta and helicities of the external massless particles are denoted by $p_i$ and $h_i$, with $p_i$ written simply as the index~$i$, with $i=1,\ldots,n$. The convention here is that all momenta are outgoing.

From the BCFW recursion relations we know that any tree amplitude is given by a sum of subamplitudes each factorizing into lower point amplitudes. Recursively the tree amplitude
can be expressed in terms of {\em primitive} three-point amplitudes. This requires in every factorizing recursion step to {\em deform} the momenta of the external particles,
\begin{equation}
\hat{p}_i = p_i + z r_i
\end{equation}
with complex momenta $r_i$ and one common complex number~$z$.
 This analytic continuation of the external momenta is done such that
on-shellness as well as momentum conservation are maintained. In general the recursion relations 
may give boundary terms. 
Here we shall consider only theories where the boundary terms do not appear. As has been shown this is the case for gauge theories as well as gravity~\cite{Arkani-Hamed:2008bsc, Feng:2009ei}.

As an example we may think of a QCD tree amplitude. Since it is an on-shell tree amplitude it is trivalent. Let us remind ourselves
that it is a peculiarity of Feynman diagrams that we have to consider 
the four-gluon vertex in QCD. 
Any $n$-point trivalent tree amplitude, with $n \ge 3$ has the 
following number of couplings, propagators, and subamplitudes,
\begin{alignat}{4} \label{numamp}
& \text{ext. particles} \quad &&
\text{couplings} \quad &&
\text{propagators} \quad &&
\text{subamplitudes} \nonumber \\
& n && n\!-\!2 && n\!-\!3 && (2n\!-\!5)!!
\end{alignat}
We have as usual denoted by
$1!!=1$,  $3!!=3 \cdot 1$, $5!!=5 \cdot 3 \cdot 1$, $\ldots$.
The number of couplings in any trivalent subamplitude equals the number of  
{\em primitive} three-point amplitudes from which it is built.
An example is the $n=6$ gluon amplitude, say with all-plus helicities, consisting of $(2n-5)!!=105$ trivalent subamplitudes, one of them is\\
\begin{tikzpicture}[scale=0.7]
\draw (0,0) -- (1,0);
\draw (1,0) -- (1,1);
\filldraw[black] (0,0) node[anchor=east] {$1^+$};
\filldraw[black] (1,0) circle (2pt);
\draw (1,0) -- (2,0);
\filldraw[black] (2,0) circle (2pt);
\draw (2,0) -- (2,1);
\filldraw[black] (2,1) node[anchor=south] {$4^+$};
\draw (2,0) -- (3,0);
\filldraw[black] (3,0) circle (2pt);
\draw (3,0) -- (3,1);
\filldraw[black] (3,1) node[anchor=south] {$5^+$};
\draw (3,0) -- (4,0);
\filldraw[black] (3,0) circle (2pt);
\draw (3,0) -- (3,1);
\filldraw[black] (4,0) node[anchor=west] {$6^+$};
\filldraw[black] (1,1) circle (2pt);
\draw (1,1) -- (0,2);
\draw (1,1) -- (2,2);
\filldraw[black] (0,2) node[anchor=south east] {$2^+$};
\filldraw[black] (2,2) node[anchor=south west] {$3^+$};
\end{tikzpicture}
\hfill \eqnum\label{ex6g}

Suppose that the particles obey an underlying 
color structure, that is, they fulfill the Jacobi identity. Then we know
from the color-kinematics relations
\cite{DelDuca:1999rs} that we can write the trivalent subamplitude in terms of subamplitudes of {\em multi-peripheral} form or equivalently denoted by {\em half-ladder} form. 
The subamplitude~\eqref{ex6g} can then be written as minus the sum of the two subamplitudes,\\
\begin{tikzpicture}[scale=0.7]
\draw (0,0) -- (1,0);
\draw (1,0) -- (1,1);
\filldraw[black] (1,1) node[anchor=south] {$2^+$};
\filldraw[black] (0,0) node[anchor=east] {$1^+$};
\filldraw[black] (1,0) circle (2pt);
\draw (1,0) -- (2,0);
\filldraw[black] (2,0) circle (2pt);
\draw (2,0) -- (2,1);
\filldraw[black] (2,1) node[anchor=south] {$3^+$};
\draw (2,0) -- (3,0);
\filldraw[black] (3,0) circle (2pt);
\draw (3,0) -- (3,1);
\filldraw[black] (3,1) node[anchor=south] {$4^+$};
\draw (3,0) -- (4,0);
\filldraw[black] (4,0) circle (2pt);
\draw (4,0) -- (4,1);
\filldraw[black] (4,1) node[anchor=south] {$5^+$};
\draw (4,0) -- (5,0);
\filldraw[black] (4,0) circle (2pt);
\draw (4,0) -- (4,1);
\filldraw[black] (5,0) node[anchor=west] {$6^+$};
\end{tikzpicture}
\hfill\\
\begin{tikzpicture}[scale=0.7]
\draw (0,0) -- (1,0);
\draw (1,0) -- (2,1);
\filldraw[black] (1,1) node[anchor=south] {$2^+$};
\filldraw[black] (0,0) node[anchor=east] {$1^+$};
\filldraw[black] (1,0) circle (2pt);
\draw (1,0) -- (2,0);
\filldraw[black] (2,0) circle (2pt);
\draw (2,0) -- (1,1);
\filldraw[black] (2,1) node[anchor=south] {$3^+$};
\draw (2,0) -- (3,0);
\filldraw[black] (3,0) circle (2pt);
\draw (3,0) -- (3,1);
\filldraw[black] (3,1) node[anchor=south] {$4^+$};
\draw (3,0) -- (4,0);
\filldraw[black] (4,0) circle (2pt);
\draw (4,0) -- (4,1);
\filldraw[black] (4,1) node[anchor=south] {$5^+$};
\draw (4,0) -- (5,0);
\filldraw[black] (4,0) circle (2pt);
\draw (4,0) -- (4,1);
\filldraw[black] (5,0) node[anchor=west] {$6^+$};
\end{tikzpicture}
\hfill\eqnum\label{ex6gc}

But for the moment we keep the amplitudes completely general, 
that is, we do not assume that they are decomposed into multi-peripheral form.\\

What about the three-point primitive amplitudes from which all tree amplitudes are formed?
The kinematics of these primitive amplitudes is completely determined by little-group scaling, that is Lorentz invariance. 
An interaction of three scalars can not have any dependence on the kinematic variables, 
that is, external momenta. For particles with any helicity $h_i$, $i=1,2,3$, with the convention that all particles are outgoing, little-group scaling gives
the {\em primitive} amplitudes (see for instance the review~\cite{Elvang:2013cua})
\begin{multline} \label{primitive}
A_3(1^{h_1}, 2^{h_2}, 3^{h_3}) = \\
\begin{cases}
 g [12]^{h_1+h_2-h_3} [23]^{h_2+h_3-h_1} [31]^{h_3+h_1-h_2},\\
g \langle 12\rangle^{h_3-h_1-h_2}  \langle 23\rangle^{h_1-h_2-h_3} \langle 31\rangle^{h_2-h_3-h_1}\;
\end{cases}
\end{multline}
with the above expression for $h_1+h_2+h_3 \ge 0$ and the lower one for
$ h_1+h_2+h_3<0$.
Here $g$ is a generic coupling constant which is a free parameter. Also the color structure is not fixed and
depends on the considered gauge group. Of course, momentum conservation and 
on-shellness of the three particles are in general not compatible and therefore, this three-point amplitudes vanish for massless particles - except for soft or collinear configurations. 
Note that the BCFW recursion relations may give factorizations into $A_3$ amplitudes, 
however with deformed momenta. For this reason theses amplitudes may give non-vanishing contributions.
For instance, the four-gluon amplitude $A_4$ can be expressed in terms of products of two primitive $A_3$ amplitudes with deformed momenta. 
 
The four-scalar~$\phi$ interaction, as known for instance from the Higgs self interaction in the Standard Model, is not constructible via the BCFW recursion relations. 
However a primitive scalar amplitude of two massless scalars $\phi$ and one massive scalar $\chi$ with mass $m_\chi$, that is, $A_3(1_\phi, 2_\phi, 3_\chi) = g$ with generic coupling $g$ provides
 a four-point amplitude $A_4(1_\phi, 2_\phi, 3_\phi, 4_\phi)$ for the massless scalars  in terms of three trivalent subamplitudes~\cite{Benincasa:2007xk}:\\
\begin{tikzpicture}[scale=0.7]
\draw[dashed] (0,0) -- (1,1);
\filldraw[black] (0,0) node[anchor=east] {1};
\draw[dashed] (0,2) -- (1,1);
\filldraw[black] (1,1) circle (2pt);
\filldraw[black] (0,2) node[anchor=east] {2};
\draw[dashed,thick] (1,1) -- (3,1);
\filldraw[black] (2,1) node[anchor=north] {$\frac{1}{s-m_\chi^2}$};
\filldraw[black] (3,1) circle (2pt);
\draw[dashed] (3,1) -- (4,0);
\filldraw[black] (4,0) node[anchor=west] {3};
\draw[dashed] (3,1) -- (4,2);
\filldraw[black] (4,2) node[anchor=west] {4};
\end{tikzpicture}\\
\begin{tikzpicture}[scale=0.7]
\draw[dashed] (0,0) -- (1,1);
\filldraw[black] (0,0) node[anchor=east] {1};
\draw[dashed] (0,2) -- (1,1);
\filldraw[black] (1,1) circle (2pt);
\filldraw[black] (0,2) node[anchor=east] {3};
\draw[dashed,thick] (1,1) -- (3,1);
\filldraw[black] (2,1) node[anchor=north] {$\frac{1}{t-m_\chi^2}$};
\filldraw[black] (3,1) circle (2pt);
\draw[dashed] (3,1) -- (4,0);
\filldraw[black] (4,0) node[anchor=west] {2};
\draw[dashed] (3,1) -- (4,2);
\filldraw[black] (4,2) node[anchor=west] {4};
\end{tikzpicture}\\
\begin{tikzpicture}[scale=0.7]
\draw[dashed] (0,0) -- (1,1);
\filldraw[black] (0,0) node[anchor=east] {1};
\draw[dashed] (0,2) -- (1,1);
\filldraw[black] (1,1) circle (2pt);
\filldraw[black] (0,2) node[anchor=east] {4};
\draw[dashed,thick] (1,1) -- (3,1);
\filldraw[black] (2,1) node[anchor=north] {$\frac{1}{u-m_\chi^2}$};
\filldraw[black] (3,1) circle (2pt);
\draw[dashed] (3,1) -- (4,0);
\filldraw[black] (4,0) node[anchor=west] {3};
\draw[dashed] (3,1) -- (4,2);
\filldraw[black] (4,2) node[anchor=west] {2};
\end{tikzpicture}
\hfill\eqnum\label{4pointtree}\\
As usual, the Mandelstam variables are denoted by $s=(p_1+p_2)^2$, $t=(p_1+p_3)^2$, and $u=(p_1+p_4)^2$. For large $g$ and large $m_\chi$, keeping 
$-g^2/m_\chi^2 \equiv \lambda$ constant, we get the desired
four point amplitude for the massless scalar $\phi$.\\

We can reverse the argument of the BCFW recursion relations and
 glue together any tree amplitude by primitive amplitudes. Following 
 BCFW we have to glue together trivalent subamplitudes in all possible ways
in order to represent a physical amplitude. The BCFW recursion relations deform the 
momenta in such a way that momentum conservation and onshellness are maintained. 
Therefore, reversing the argument of BCFW by glueing together primitive $A_3$ amplitudes in order to form a given amplitude $A_n$ we can never form a loop.
This is clear because for given external momenta a loop inevitably violates onshellness. Virtual particles, appearing in Feynman diagrams, are absent in amplitudes.

\section{Tree amplitudes forming higher orders}

Suppose we want to calculate the amplitude $A_n(1^{h_1}, \ldots, n^{h_n})$.
Following BCFW we have to glue together primitive amplitudes $A_3$ in all possible ways forming this amplitude.
In order to do this we have to deform the momenta in general, complying with BCFW. However, the amplitude
$A_n(1^{h_1}, \ldots, n^{h_n})$ is indistinguishable from amplitudes with additional pairs of particles in the forward limit: 
For instance, adding one pair of particles with opposite momenta and opposite helicities, the amplitude $A_{n+2}^{\text{1 pair}}(1^{h_1}, \ldots, n^{h_n}, - l_1^{-h_{l_1}}, l_1^{h_{l_1}})$ is indistinguishable from the amplitude
$A_n(1^{h_1}, \ldots, n^{h_n})$. The pair of additional particles are balanced and together form a vacuum state. 
In general we find that along with the amplitude $A_n(1^{h_1}, \ldots, n^{h_n})$
we have to take into account the amplitudes with $i = 1, 2, 3, \ldots, k$ additional pairs of particles in the forward limit:
\begin{equation} \label{Akpairs}
A_{n+2k}^{k \text{ pairs}}(1^{h_1}, \ldots, n^{h_n}, - l_1^{-h_{l_1}}, l_1^{h_{l_1}}, \ldots, - l_k^{-h_{l_k}}, l_k^{h_{l_k}}) .
\end{equation}
With every additional pair we add two primitive $A_3$ amplitudes, increase the number of couplings
by two, that is, every additional pair in the forward limit increases 
the order by one unit. 
The forward momenta $-l_i$ and $l_i$, $i=1,\ldots,k$ are opposite but arbitrary,
therefore we have to integrate over their phase space.
For the amplitude with $k$ additional pairs in the forward limit we get
\begin{multline} \label{ampfor1}
(-1)^{2 (h_{l_1} + \ldots + h_{l_k})} \times \\
\int\limits_{l_1, \ldots, l_k} 
A_{n+2k}^{k \text{ pairs}}(1^{h_1}, \ldots, n^{h_n}, - l_1^{-h_{l_1}}, l_1^{h_{l_1}}, \ldots, - l_k^{-h_{l_k}}, l_k^{h_{l_k}}),
\end{multline}
where we abbreviate the integrations as
\begin{equation}
\int\limits_{l_i} = \int \frac{ d^D l_i}{(2\pi)^D} 2\pi \delta^{(+)}(l_i).
\end{equation} 
As usual we write $\delta^{(+)}(l_i) = \theta(l_{i0}) \delta(l_i^2)$. We have to perform the calculation in general $D=4 - 2 \epsilon$  dimensions in order to regularize the in general singular integrals. The prefactor in~\eqref{ampfor1}
gives a factor of $-1$ for every additional fermion pair.

We emphasize that all the subamplitudes which appear on the right-hand side of
\eqref{ampfor1} with $k$ additional pairs are trivalent or, as we will see, products of trivalent amplitudes.
Any loop of on-shell amplitudes would violate onshellness. 
In the expression~\eqref{ampfor1} the number of additional pairs $k$ corresponds to subamplitudes at $\text{N}^k\text{LO}$. 
However, we have to take into account 
also split - or cut - subamplitudes, where propagators are replaced by further pairs in the forward limit; without increasing the perturbation order. Note that the order of the amplitudes is defined here with respect to the number of trivalent couplings~\eqref{numamp}. A leading order $n$-point amplitude therefore corresponds to $n-2$ trivalent couplings, irrespective whether it is vanishing or not.  

Let us consider as an example the six-gluon maximally helicity violating amplitude, that is two gluons carry negative, all remaining ones carry plus helicity. One of the trivalent subamplitudes at leading order (LO) of $A_6(1^-,2^-,3^+,4^+,5^+,6^+)$ is\\
\begin{tikzpicture}[scale=0.7]
\draw (0,0) -- (1,0);
\draw (1,0) -- (1,1);
\filldraw[black] (0,0) node[anchor=east] {$1^-$};
\filldraw[black] (1,1) node[anchor=south] {$2^-$};
\filldraw[black] (1,0) circle (2pt);
\draw (1,0) -- (2,0);
\draw (2,0) -- (2,1);
\filldraw[black] (2,1) node[anchor=south] {$3^+$};
\filldraw[black] (2,0) circle (2pt);
\draw (2,0) -- (3,0);
\draw (3,0) -- (3,1);
\filldraw[black] (3,1) node[anchor=south] {$4^+$};
\filldraw[black] (3,0) circle (2pt);
\draw (3,0) -- (4,0);
\draw (4,0) -- (4,1);
\filldraw[black] (4,1) node[anchor=south] {$5^+$};
\filldraw[black] (4,0) circle (2pt);
\draw (4,0) -- (5,0);
\filldraw[black] (5,0) node[anchor=west] {$6^+$};
\end{tikzpicture}\hfill\eqnum\label{exA6LO}\\
Obviously, we can not split this leading-order amplitude by replacing 
a propagator by a vacuum pair, for instance,\\
\begin{tikzpicture}[scale=0.7]
\draw (0,0) -- (1,0);
\draw (1,0) -- (1,1);
\filldraw[black] (0,0) node[anchor=east] {$1^-$};
\filldraw[black] (1,1) node[anchor=south] {$2^-$};
\filldraw[black] (1,0) circle (2pt);
\draw (1,0) -- (2,0);
\draw (2,0) -- (2,1);
\filldraw[black] (2,1) node[anchor=south] {$3^+$};
\filldraw[black] (2,0) circle (2pt);
\draw (2,0) -- (3,0);
\filldraw[black] (3,0) node[anchor=west] {$-l_1^-$};
\filldraw[black] (6,0) node[anchor=east] {$l_1^+$};
\draw (6,0) -- (7,0);
\draw (7,0) -- (7,1);
\filldraw[black] (7,1) node[anchor=south] {$4^+$};
\filldraw[black] (7,0) circle (2pt);
\draw (7,0) -- (8,0);
\draw (8,0) -- (8,1);
\filldraw[black] (8,1) node[anchor=south] {$5^+$};
\filldraw[black] (8,0) circle (2pt);
\draw (8,0) -- (9,0);
\filldraw[black] (9,0) node[anchor=west] {$6^+$};
\end{tikzpicture}\hfill\eqnum\label{exA6LOsplit}\\
This is not possible since we have to satisfy both,
on-shellness and momentum conservation but 
$l_1^2= (p_1+p_2+p_3)^2 \neq 0$ for general external momenta.

Now let us go beyond LO and add one pair in the forward limit;
 a subamplitude of 
$A_{6+2}^{\text{1 pair}}(1^-,2^-,3^+,4^+,5^+,6^+,-l_1^-,l_1^+)$ is\\
\begin{tikzpicture}[scale=0.7]
\draw (0,0) -- (1,0);
\draw (1,0) -- (1,1);
\filldraw[black] (0,0) node[anchor=east] {$1^-$};
\filldraw[black] (1,1) node[anchor=south] {$2^-$};
\filldraw[black] (1,0) circle (2pt);
\draw (1,0) -- (2,0);
\draw (2,0) -- (2,1);
\filldraw[black] (2,1) node[anchor=south] {$3^+$};
\filldraw[black] (2,0) circle (2pt);
\draw (2,0) -- (3,0);
\draw (3,0) -- (3,1);
\filldraw[black] (3,1) node[anchor=south] {$-l_1^-$};
\filldraw[black] (3,0) circle (2pt);
\draw (3,0) -- (4,0);
\draw (4,0) -- (4,1);
\filldraw[black] (4,1) node[anchor=south] {$4^+$};
\filldraw[black] (4,0) circle (2pt);
\draw (4,0) -- (5,0);
\draw (5,0) -- (5,1);
\filldraw[black] (5,1) node[anchor=south] {$5^+$};
\filldraw[black] (5,0) circle (2pt);
\draw (5,0) -- (6,0);
\draw (6,0) -- (6,1);
\filldraw[black] (6,1) node[anchor=south] {$l_1^+$};
\filldraw[black] (6,0) circle (2pt);
\draw (6,0) -- (7,0);
\filldraw[black] (7,0) node[anchor=west] {$6^+$};
\end{tikzpicture}\hfill\eqnum\label{exA6NLO}\\
The number of couplings is increased by two, that is  $k=1$, corresponding to
an amplitude at NLO. At the same order we have also 
to consider the split subamplitude\\
$A_{6+2+2}^{\text{1 pair, 1 split}}(1^-,2^-,3^+,4^+,5^+,6^+,-l_1^-,l_1^+,-l_2^-,l_2^+)$, for instance,\\
\begin{tikzpicture}[scale=0.7]
\draw (0,0) -- (1,0);
\draw (1,0) -- (1,1);
\filldraw[black] (0,0) node[anchor=east] {$1^-$};
\filldraw[black] (1,1) node[anchor=south] {$2^-$};
\filldraw[black] (1,0) circle (2pt);
\draw (1,0) -- (2,0);
\draw (2,0) -- (2,1);
\filldraw[black] (2,1) node[anchor=south] {$3^+$};
\filldraw[black] (2,0) circle (2pt);
\draw (2,0) -- (3,0);
\draw (3,0) -- (3,1);
\filldraw[black] (3,1) node[anchor=south] {$-l_1^-$};
\filldraw[black] (3,0) circle (2pt);
\draw (3,0) -- (4,0);
\draw (4,0) -- (4,1);
\filldraw[black] (4,1) node[anchor=south] {$4^+$};
\filldraw[black] (4,0) circle (2pt);
\draw (4,0) -- (5,0);
\filldraw[black] (5,0) node[anchor=west] {$-l_2^-$};
\filldraw[black] (8,0) node[anchor=east] {$l_2^+$};
\draw (8,0) -- (9,0);
\draw (9,0) -- (9,1);
\filldraw[black] (9,1) node[anchor=south] {$5^+$};
\filldraw[black] (9,0) circle (2pt);
\draw (9,0) -- (10,0);
\draw (10,0) -- (10,1);
\filldraw[black] (10,1) node[anchor=south] {$l_1^+$};
\filldraw[black] (10,0) circle (2pt);
\draw (10,0) -- (11,0);
\filldraw[black] (11,0) node[anchor=west] {$6^+$};
\end{tikzpicture}
\phantom{space} \hfill\eqnum\label{exA6NLOs}\\
This amplitude in the form of a product of two {\em partial} amplitudes, denoted by the superscript ‘‘1 split'', 
does neither violate momentum conservation nor on-shellness. It is also physically indistinguishable from the 
leading order tree amplitude.

Alternatively we may replace different propagators  - as long as we do not split into a vanishing $A_3$ subamplitude. We see that at NLO we can at most replace one propagator, that is split once into two partial amplitudes, respecting on-shellness and momentum conservation. 

For an arbitrary order $k$, that is, amplitudes at $\text{N}^k\text{LO}$, we have to add $k$ pairs of external forward particles. 
After the addition of these $k$ pairs of external particles we can, without 
changing the order $k$, cut propagators, that is, replace propagators by
additional pairs in the forward limit. By every such cut of a propagator we split the amplitude into separate parts. 
Obviously there is an upper limit of
propagators we can replace:
 The maximal number of propagators we may replace by particles in the forward limit is equal to the number of pairs of additional external particles~$k$. 
This comes from the fact that 
we have to fulfill simultaneously on-shellness and momentum conservation.
We get therefore at a certain perturbation order $k$ exactly $k$ pairs of external particles in the forward limit with afterwards $0,\ldots,k$ replaced propagators. The number of parts into which the subamplitudes split is given by the number of replaced propagators plus one. We thus get the composition of tree sub and partial amplitudes at order $\text{N}^k\text{LO}$:
\begin{multline} \label{ampfor}
A_n^{\text{N}^k\text{LO}} (1^{h_1}, \ldots, n^{h_n}) =\\
(-1)^{2 (h_{l_1} + \ldots + h_{l_k})} \times\\
\int\limits_{l_1, \ldots, l_k} 
A_{n+2k}^{k \text{ pairs}}(1^{h_1}, \ldots, n^{h_n}, - l_1^{-h_{l_1}}, l_1^{h_{l_1}}, \ldots, - l_k^{-h_{l_k}}, l_k^{h_{l_k}})
\end{multline}
with
\begin{multline*} 
A_{n+2k}^{k \text{ pairs}}
=  
A_{n+2k}^{k \text{ pairs}, 0 \text{ split}}
+ 
(-1)^{2 h'_1} \int\limits_{l'_1} A_{n+2k+2}^{k \text{ pairs}, 1 \text{ split}}
+ 
\ldots\\
+
(-1)^{2(h'_1+\ldots h'_k)} \int\limits_{l'_1,\ldots,l'_k} A_{n+2k+2k}^{k \text{ pairs}, k \text{ splits}}
\end{multline*}
and partial amplitudes~$i=1,\ldots,k$,
\begin{multline*}
A_{n+2k+2i}^{k \text{ pairs}, i \text{ splits}}
(1^{h_1}, \ldots, n^{h_n}, \!
- l_1^{-h_{l_1}}\!, l_1^{h_{l_1}}\!, \ldots,\!
- l_k^{-h_{l_k}}\!, l_k^{h_{l_k}}\!, 
\\
 {-l'}_1^{-h'_{1}}, {l'}_1^{h'_{1}}, \ldots, - {l'}_i^{-h'_{i}} , {l'}_i^{h'_{i}}
).
\end{multline*}
Similar to the  pairs added to the trivalent amplitude we have to 
 integrate over the phase space of the momenta ($l'_i$) of the cut propagators. 
Let us illustrate this in an example - the all-plus four-gluon amplitude. The leading order ($k=0$) amplitude with $n=4$, that is, $A_4$ consists of $(2n-5)!!=3$ subamplitudes with coupling order $n-2=2$, that is, $g^2$. At the next-to-leading order ($k=1$), we have to add one external pair of forward particles
in order to increase the number of couplings by two. For $k=1$ we can either cut no propagator at all or at most cut one propagator in the resulting subamplitude.
 Without any propagator replacements we get one trivalent amplitude and with one cut we get the subamplitude split into two. 
At next-to-next-to leading order, that is, $\text{N}^k\text{L0}$ with
$k=2$,  we have to add
exactly two pairs of forward particles. In these
amplitudes we may either cut no propagator or cut one or maximally two, 
yielding trivalent subamplitudes unsplit, split into two or into three parts.\\

With~\eqref{ampfor} we have written the amplitudes to any perturbation order
in terms of trivalent subamplitudes or products of trivalent partial amplitudes. 
We have seen that we get split amplitudes,
that is, products of tree amplitudes, where the number of maximal splittings equals the perturbation order. Eventually, the amplitude can be expressed in terms of tree amplitudes and we can apply the BCFW recursion relations. This means that we can express scattering amplitudes in terms of primitive $A_3$ amplitudes. From the splittings we get the different color structures: By color kinematics every trivalent subamplitude gives a trace of color generators and we see that we get the known products of traces for higher order scattering amplitudes.

\section{Example four-gluon amplitude}

In this section we want to show how the four-gluon interaction 
is determined in terms of tree amplitudes. With respect to an increasing perturbation order we have to add systematically, as outlined in the previous sections, particle pairs in the forward limit. 

The four-gluon amplitude provides an excellent framework to study the methods.
It is the lowest-point amplitude complying with on-shellness and momentum conservation for massless particles. The simplest one is the all-plus helicity amplitude, but all remaining ones 
can be studied in a similar way.\\

At leading order, the amplitude vanishes for all plus helicities for any number of external gluons,
\begin{equation} \label{allplus0}
A_n^{\text{LO}}(1^+, 2^+, \ldots, n^+) =0\;.
\end{equation}
The usual proof is given with the help of polarization ‘‘tensors''. Not resorting
to polarization ‘‘tensors''
we can easily see that the partial amplitudes~\eqref{allplus0}
have to vanish from the recursive application of BCFW.
First of all we can assume that the amplitudes are in multi-peripheral form by the application of color-kinematics relations~\cite{DelDuca:1999rs}. 
By a convenient relabeling of the external momenta we can apply BCFW as indicated in the following amplitude,\\
\begin{tikzpicture}[scale=0.7]
\draw (0,0) -- (1,0);
\draw (1,0) -- (1,1);
\filldraw[black] (0,0) node[anchor=east] {$1^+$};
\filldraw[black] (1,1) node[anchor=south] {$\hat{2}^+$};
\filldraw[black] (1,0) circle (2pt);
\draw (1,0) -- (2,0);
\filldraw[black] (1,0) node[anchor=north west] {$-$};
\filldraw[black] (3,0) node[anchor=north east] {$+$};
\draw (2,0) -- (3,0);
\filldraw[black] (2,0) node[anchor=south] {$\hat{P}_{12}$};
\filldraw[black] (3,0) circle (2pt);
\draw (3,0) -- (3,1);
\filldraw[black] (3,1) node[anchor=south] {$\hat{3}^+$};
\draw (3,0) -- (4,0);
\filldraw[black] (4,0) node[anchor=west] {$\cdots$};
\draw (5,0) -- (6,0);
\draw (6,0) -- (6,1);
\filldraw[black] (6,0) circle (2pt);
\filldraw[black] (6,1) node[anchor=south] {$(n-1)^+$};
\draw (6,0) -- (7,0);
\filldraw[black] (7,0) node[anchor=west] {$n^+$};
\end{tikzpicture}
\hfill\eqnum\label{Ang+}\\
We choose
a $|2, 3 \rangle$ shift with 
the helicities as indicated and the unshifted propagator given by $P_{12} = 1/(p_1+p_2)^2$. The opposite choice
of helicities for the particles connected to the propagator $\hat{P}_{12}$ is excluded because it is directly proportional to a vanishing all-plus amplitude $A_3$. However with further recursions of BCFW we can not avoid to get one vanishing all-plus amplitude $A_3$ and therefore the all-plus $n$-gluon amplitude vanishes at leading order.

The situation changes if we consider the amplitude beyond leading
order, employing~\eqref{ampfor}, for instance at next-to-leading oder we have to consider  subamplitudes of the form $A_{n+2}^{\text{1 pair}}(1, \ldots, n, -l_1^-, l_1^+)$ with a pair of gluons added in the forward limit. 
Applying color-kinematics relations, we get the subamplitudes in multi-peripheral form,  
where we have the extra pair of external particles with momentum, respectively helicity $-l_1^-$, and $l_1^+$ in all possible positions. Inevitably, we encounter now one external gluon with negative helicity. Then, in the chain of $A_3$ primitive amplitudes, similar to~\eqref{Ang+} we can choose the helicities of the internal lines
such that there never appears an all-plus amplitude $A_3$ and the argument above does not apply.

In supersymmetry, we immediately see that the all-plus amplitudes have to vanish at all orders, because for every external additional pair of bosons, at any position, we have to add an amplitude where the bosons are replaced by fermions. The corresponding factor $(-1)^{2 h_{l_i}}$ in~\eqref{ampfor} ensures that both contributions enter with opposite signs and therefore the sum of partial amplitudes to all orders of 
all-plus (or equivalentely all minus) amplitudes vanishes in supersymmetry.

\subsection{NLO all-plus four-gluon amplitude}

The conventional method is to compute the amplitude via Feynman diagrams. An alternative approach~\cite{Bern:1994zx,Giele:2008ve} is based on unitary cuts
applied to an expansion of the amplitude in terms of scalar integrals. 
In~\cite{Maniatis:2019pig} the all-plus four-gluon amplitude has been calculated employing the Feynman-tree theorem. Here we want to show the method in a changed manner explicitly starting from primitive $A_3$ amplitudes integrating over additional pairs of particles in the forward limit.

As has been argued before, the factor $(-1)$ in~\eqref{ampfor} ensures that the $n$-gluon amplitude vanishes to all orders in supersymmetry (SYM), in particular in ${\cal N}=1$ and ${\cal N}=4$ SYM. 
From the vanishing amplitude, a simple context between the additional pairs of bosons and fermions can be deduced~\cite{Bern:1996ja}:
In a supersymmetric $n$-gluon amplitude we have to consider
additional pairs of scalars, fermions and gluons in general. 
Here we want to calculate the next-to-leading order amplitude
so we have to consider one additional pair of forward particles. In ${\cal N}=4$ SYM
we have at next-to-leading order for the vanishing amplitude one additional
pair of gluons, $(g)$, paired with 4 fermions, $(f)$, and 3 complex scalars, $(s)$:
\begin{equation} \label{AnN4}
A_{n+2}^{{\cal N}=4, \text{ 1 pair}}=
A_{n+2}^{\text{1 pair } (g)} -
4\;A_{n+2}^{\text{1 pair } (f)}+
3\;A_{n+2}^{\text{1 pair } (s)}\;.
\end{equation}
In ${\cal N}=1$ SYM
we have at next-to-leading order for the vanishing amplitude one additional
pair of gluons, $(g)$, paired with 1 fermion, $(f)$:
\begin{equation}  \label{AnN1}
A_{n+2}^{{\cal N}=1, \text{ 1 pair}}=
A_{n+2}^{\text{1 pair } (g)} -
A_{n+2}^{\text{ 1 pair } (f)}\;.
\end{equation}
Since the amplitudes~\eqref{AnN4} and \eqref{AnN1} vanish
for all-plus helicity gluons we see that
\begin{equation}
A_{n+2}^{\text{1 pair } (g)} = A_{n+2}^{\text{1 pair } (s)}\;.
\end{equation}

The task now is to calculate the amplitude~\eqref{ampfor}
with one additional pair of scalars in the forward limit:
\begin{multline}
A_4^{\text{NLO}}(1^+, 2^+, 3^+, 4^+) =\\
\int\limits_{l_1} A_{4+2}^{\text{1 pair } (s)}(1^+, 2^+, 3^+, 4^+, -l_1, l_1).
\end{multline}
The subamplitude $A_{4+2}^{\text{1 pair} (s)}$ has one contribution
without splitting and one partial amplitud split once~\eqref{ampfor}, where we suppress
from now on the superscript $(s)$:
\begin{multline} \label{A4split}
A_{4+2}^{\text{1 pair }}(1^+, 2^+, 3^+, 4^+, -l_1, l_1) =\\
A_{4+2}^{\text{1 pair, 0 splits }}(1^+, 2^+, 3^+, 4^+, -l_1, l_1) \\
+
\int\limits_{l_2}
A_{4+2+2}^{\text{1 pair, 1 split }}(1^+, 2^+, 3^+, 4^+, -l_1, l_1, -l_2, l_2) \;.
\end{multline}
Eventually we find the expression \\
\begin{multline} \label{A4dec1}
A_4^{\text{NLO}}(1^+, 2^+, 3^+, 4^+)  
\\=
\int\limits_{l_1} 
A_{4+2}^{\text{1 pair, 0 splits }}(1^+, 2^+, 3^+, 4^+, -l_1, l_1) \\
+
\int\limits_{l_1, l_2}
A_{4+2+2}^{\text{1 pair, 1 split }}(1^+, 2^+, 3^+, 4^+, -l_1, l_1, -l_2, l_2)
\\ \equiv
A_{4}^{\text{1 pair, 0 splits }}+A_{4}^{\text{1 pair, 1 split }} .
\end{multline}
A subamplitude of the first term on the r.h.s. of ~\eqref{A4dec1} can be graphically represented as\\
\begin{tikzpicture}[scale=0.7]
\draw[dashed] (0,0) -- (1,0);
\draw (1,0) -- (1,1);
\filldraw[black] (0,0) node[anchor=east] {$-l_1$};
\filldraw[black] (1,1) node[anchor=south] {$1^+$};
\filldraw[black] (1,0) circle (2pt);
\draw[dashed] (1,0) -- (2,0);
\draw (2,0) -- (2,1);
\filldraw[black] (2,1) node[anchor=south] {$2^+$};
\filldraw[black] (2,0) circle (2pt);
\draw[dashed] (2,0) -- (3,0);
\draw (3,0) -- (3,1);
\filldraw[black] (3,1) node[anchor=south] {$3^+$};
\filldraw[black] (3,0) circle (2pt);
\draw[dashed] (3,0) -- (4,0);
\draw (4,0) -- (4,1);
\filldraw[black] (4,1) node[anchor=south] {$4^+$};
\filldraw[black] (4,0) circle (2pt);
\draw[dashed] (4,0) -- (5,0);
\filldraw[black] (5,0) node[anchor=west] {$l_1$};
\end{tikzpicture}
 \hfill\eqnum\label{A4p6}\\
The gluon is drawn as a full line and the complex scalar as a dashed line. 
A subamplitude with one splitting, that is, with
one propagator replaced by a pair in the forward limit corresponds to the partial amplitude
 in the second term on the r.h.s. of~\eqref{A4dec1},\\
\begin{tikzpicture}[scale=0.7]
\draw[dashed] (0,0) -- (1,0);
\draw (1,0) -- (1,1);
\filldraw[black] (0,0) node[anchor=east] {$-l_1$};
\filldraw[black] (1,1) node[anchor=south] {$1^+$};
\filldraw[black] (1,0) circle (2pt);
\draw[dashed] (1,0) -- (2,0);
\draw (2,0) -- (2,1);
\filldraw[black] (2,1) node[anchor=south] {$2^+$};
\filldraw[black] (2,0) circle (2pt);
\draw[dashed] (2,0) -- (3,0);
\filldraw[black] (3,0) node[anchor=west] {$-l_2$};
\end{tikzpicture}\qquad
\begin{tikzpicture}[scale=0.7]
\draw[dashed] (0,0) -- (1,0);
\draw (1,0) -- (1,1);
\filldraw[black] (0,0) node[anchor=east] {$l_2$};
\filldraw[black] (1,1) node[anchor=south] {$3^+$};
\filldraw[black] (1,0) circle (2pt);
\draw[dashed] (1,0) -- (2,0);
\draw (2,0) -- (2,1);
\filldraw[black] (2,1) node[anchor=south] {$4^+$};
\filldraw[black] (2,0) circle (2pt);
\draw[dashed] (2,0) -- (3,0);
\filldraw[black] (3,0) node[anchor=west] {$l_1$};
\end{tikzpicture}
\hfill\eqnum\label{A4p33}\\
Note that a subamplitude with one additional pair, as shown in~\eqref{A4p6} can only be
cut in the propagator as shown in~\eqref{A4p33} since all other propagators cut yield 
an isolated (unshifted) $A_3$ amplitude. We recall that 
the primitive amplitudes $A_3$ is only non-vanishing for BCFW shifted,
that is, deformed, momenta.
Let us assume that the gluons are color ordered. Then we have for the in total $4!$ permutations of the external gluons only to take the cyclic
permutations $1 \to 2 \to 3 \to 4 \to 1$ into account.
Following~\eqref{A4dec1} we have to integrate over the phase space of the particle pairs in the forward limit.

We encounter the problem that the phase-space integration has to 
be performed in general $D$ dimensions but the
helicity formalism is defined for $4$ dimensions. 
We follow here the solution as outlined in~\cite{Bern:1991aq}: 
The $D=4-2\epsilon$ dimensional momenta $l_i$ are split into its four-dimensional and the remaining $\mu=(-2\epsilon)$ dimensional part. The external momenta are kept four dimensional. Suppose that $p$ is an external four-dimensional momentum, then we find for a propagator with momentum $(l_i-p)^2 \to (l_i-p)^2 - \mu^2$.
The integration measure changes as $d^D l_i/(2\pi)^D \to d^{-2\epsilon} \mu/(2\pi)^{-2\epsilon}  d^4 l_i/(2\pi)^4$.
We get a four-dimensional phase space integration with an artificial 
mass parameter~$\mu$ which appears in the phase space momenta in the integrand.\\

With view on~\eqref{A4dec1}, respectively, \eqref{A4p6} and \eqref{A4p33}
we have to calculate the six-point and four-point amplitudes. In this respect we follow closely~\cite{Badger:2005zh} where tree amplitudes for up to four gluons and two complex massive scalars have been computed. We start the composition of the amplitude with the
primitive $A_3$ amplitude and glue following BCFW the higher-point amplitudes together.
The primitive amplitude of one gluon with $\pm$ helicity and two massive scalars is\\
\begin{tikzpicture}[scale=0.7]
\draw[dashed] (0,0) -- (1,0);
\draw (1,0) -- (1,1);
\filldraw[black] (0,0) node[anchor=east] {$1$};
\filldraw[black] (1,1) node[anchor=south] {$2^\pm$};
\filldraw[black] (1,0) circle (2pt);
\draw[dashed] (1,0) -- (2,0);
\filldraw[black] (2,0) node[anchor=west] {$3$};
\end{tikzpicture}
\hfill\eqnum\label{A3P}\\
Since the scalar is massive, the expressions in~\eqref{primitive} have to be modified.
Depending on the helicity of the gluon, the primitive three-point amplitudes read~\cite{Badger:2005zh,Arkani-Hamed:2017jhn}
\begin{equation} \label{A3b}
A_3(1, 2^+, 3) =
\frac{ \langle q |1 | 2]} {\langle q\; 2 \rangle}\;,
\qquad
A_3(1, 2^-, 3) =
-\frac{ \langle 2 |1 | q ]} {[q\; 2]} \;.
\end{equation}
Here $q$ denotes an arbitrary null vector, linearly independent of the momenta of the particles~1 and 2.
Glueing  two primitive subamplitudes~\eqref{A3P} together,
we get the $A_4(1, 2^+,3^+,4)$ amplitude by one BCFW
step, where we shift the two massless gluon momenta, that is, a $|2,3\rangle$ shift,
\begin{equation}
|\hat{2}] = |2] + z |3], \quad
|\hat{3}] = |3], \quad
|\hat{3}\rangle = |3\rangle - z |2\rangle, \quad
|\hat{2}\rangle = |2\rangle ,
\end{equation}
\begin{tikzpicture}[scale=0.7]
\draw[dashed] (0,0) -- (1,0);
\draw (1,0) -- (1,1);
\filldraw[black] (0,0) node[anchor=east] {$1$};
\filldraw[black] (1,1) node[anchor=south] {$\hat{2}^+$};
\filldraw[black] (1,0) circle (2pt);
\draw[dashed] (1,0) -- (2,0);
\filldraw[black] (1.5,0) node[anchor=north] {$1/\hat{P}^2$};
\draw (2,0) -- (2,1);
\filldraw[black] (2,1) node[anchor=south] {$\hat{3}^+$};
\filldraw[black] (2,0) circle (2pt);
\draw[dashed] (2,0) -- (3,0);
\filldraw[black] (3,0) node[anchor=west] {$4$};
\end{tikzpicture}
\hfill\eqnum\label{A4sB}\\
and find with the propagator momentum $P^2=(p_1+p_2)^2$,
\begin{multline} \label{A4}
A_4(1,2^+, 3^+,4) = \\
A_3(\hat{2}^+, 1, -\hat{P}) \frac{1}{P^2 - \mu^2} A_3(\hat{P}, \hat{3}^+,4) =\\
- 
\frac{\langle \hat{3} | \hat{P} | \hat{2} ] } {\langle \hat{3} \hat{2} \rangle}
 \frac{1}{P^2 - \mu^2}
\frac{\langle \hat{2} | 4 | \hat{3} ] } {\langle \hat{2} \hat{3} \rangle}=
\\
\frac{\mu^2}{P^2 -\mu^2}
\frac{[23]}{\langle 23 \rangle} \;.
\end{multline}
Glueing another $A_3$ amplitude to the $A_4$ amplitude
we get the $A_5(1,2^+, 3^+,4^+,5)$ amplitude
with a $|2, 3\rangle$ shift,\\
\begin{tikzpicture}[scale=0.7]
\draw[dashed] (0,0) -- (1,0);
\draw (1,0) -- (1,1);
\filldraw[black] (0,0) node[anchor=east] {$1$};
\filldraw[black] (1,1) node[anchor=south] {$\hat{2}^+$};
\filldraw[black] (1,0) circle (2pt);
\draw[dashed] (1,0) -- (2,0);
\filldraw[black] (1.5,0) node[anchor=north] {$1/\hat{P}^2$};
\draw (2,0) -- (2,1);
\filldraw[black] (2,1) node[anchor=south] {$\hat{3}^+$};
\filldraw[black] (2,0) circle (2pt);
\draw[dashed] (2,0) -- (3,0);
\draw (3,0) -- (3,1);
\filldraw[black] (3,1) node[anchor=south] {$4^+$};
\filldraw[black] (3,0) circle (2pt);
\draw[dashed] (3,0) -- (4,0);
\filldraw[black] (4,0) node[anchor=west] {$5$};
\end{tikzpicture}
\hfill\eqnum\label{A5}

\begin{multline} \label{A5split}
A_{5}(1, 2^+, 3^+, 4^+, 5)
 =\\
 A_{3}(1, \hat{2}^+, \hat{P}) 
 \frac{1}{P^2-\mu^2}
A_{4}(-\hat{P}, \hat{3}^+,4^+,5).
\end{multline}
Here we plug in our results for $A_3$, \eqref{A3b}, and $A_4$, \eqref{A4}.
Similar we glue together via BCFW the six point amplitude
\begin{multline} \label{A6split}
A_{6}(-l_1, 1^+, 2^+, 3^+, 4^+,l_1) =\\
A_{3}(-l_1, \hat{1}^+, -\hat{P})
\frac{1}{P^2-\mu^2}
A_{5}(\hat{P}, \hat{2}^+, 3^+, 4^+, l_1) .
\end{multline}
Only one factorization is possible employing a $[1,2\rangle$ shift.
We eventually find
\begin{multline} \label{A6}
A_{6}(-l_1, 1^+, 2^+, 3^+, 4^+,l_1)
= \mu^4 \cdot \\
\frac{1}{ [(p_1-l_1)^2- \mu^2]} 
\frac{1}{[(p_1+p_2-l_1)^2- \mu^2]} \cdot
\\
\frac{1}{[(p_1+p_2+p_3-l_1)^2- \mu^2]}
\frac{ [1 2] [3 4] }
{\langle 1 2 \rangle \langle 3 4 \rangle} .
\end{multline}
With these preparations we get for the first term on the r.h.s of~\eqref{A4dec1} with~\eqref{A6}
\begin{multline} \label{A4res}
A_{4}^{\text{1 pair, 0 splits}} =
\frac{ [1 2] [3 4] }{ \langle 1 2 \rangle \langle 3 4 \rangle} 
\int \frac{d^{-2 \epsilon} \mu}{(2\pi)^{-2\epsilon}}
\mu^4 \cdot
\\
\int \frac{ d^4 l_1}{ (2\pi)^3 }
\bigg\{
\delta^{(+)}(l_1^2-\mu^2) 
\cdot\\
\frac{1}{ [(p_1-l_1)^2\!-\! \mu^2]}
\frac{1}{[(p_1+p_2-l_1)^2\!-\! \mu^2]} 
\cdot\\
\frac{1}{[(p_1+p_2+p_3-l_1)^2\!-\! \mu^2]}
\bigg\}\;.
\end{multline}
For the second term on the r.h.s. of~\eqref{A4dec1} 
with one additional pair,
split into two parts, we have
\begin{multline} \label{A41split}
A_{4}^{\text{1 pair, 1 split}}  =\\
\int \frac{d^{-2 \epsilon} \mu}{(2\pi)^{-2\epsilon}}
\mu^4
\int 	\frac{ d^4 l_1}{ (2\pi)^3 }
 		\frac{ d^4 l_2}{ (2\pi)^3 }
\delta^{(+)} (l_1^2-\mu^2) 
\delta^{(+)}(l_2^2-\mu^2)
\cdot \\
A_4(-l_1, 1^+, 2^+, -l_2) A_4(l_2, 3^+, 4^+, l_1)\;.
\end{multline}
Here we plug in our results for the two amplitudes
from~\eqref{A4} and get
\begin{multline} \label{A41s1}
A_{4}^{\text{1 pair, 1 split}} =\\
\frac{ [1 2] [3 4] }{ \langle 1 2 \rangle \langle 3 4 \rangle}
\int \frac{d^{-2 \epsilon} \mu}{(2\pi)^{-2\epsilon}}
\mu^4
\int \frac{ d^4 l_1}{ (2\pi)^3 }
 \frac{ d^4 l_2}{ (2\pi)^3 }
 \\
\frac{\delta^{(+)}(l_1^2-\mu^2)\delta^{(+)}(l_2^2-\mu^2)}
 {[(l_1-p_1)^2-\mu^2][(l_2-p_3)^2-\mu^2]} .
\end{multline}
This phase space integral is known from the Cutkosky cutting rules
and equals twice the imaginary part of a box integral which vanishes.
All other permutations of the external momenta 
give also vanishing contributions. 
Especially the kinematic factor
\begin{equation} \label{kin}
\frac{ [1 2] [3 4] }{ \langle 1 2 \rangle \langle 3 4 \rangle}
\end{equation}
is invariant under all permutations. This can be seen 
for the cyclic permutation $1 \to 2 \to 3 \to 4 \to 1$, 
using momentum conservation and 
$[12] = \langle 41 \rangle[12]/\langle 41 \rangle=
-\langle 43 \rangle [32]/\langle 41 \rangle$ and
similar $[34] = -\langle 21 \rangle [14]/\langle 23 \rangle$.
Besides, the expression~\eqref{kin} is
invariant under $1 \leftrightarrow 2$, under
$3 \leftrightarrow 4$, and under reflections, which gives invariance under all
permutations.

It remains to calculate~\eqref{A4res}.
We go conveniently
back to a phase space integral in $D$ dimensions,
\begin{multline} \label{A4D}
A_{4}^{\text{1 pair, 0 splits}} =
\frac{ [1 2] [3 4] }{ \langle 1 2 \rangle \langle 3 4 \rangle}
\cdot \\
\int \!\!\frac{ d^D l_1}{ (2\pi)^{D-1}}
\frac{\delta^{(+)}(l_1^2) }{(p_1-l_1)^2 (p_1+p_2-l_1)^2 (p_1+p_2+p_3-l_1)^2}. 
\end{multline}
The integral can be carried out directly
with the help of Schwinger parameters~\cite{Caron-Huot:2010fvq, Maniatis:2019pig}
\begin{equation} \label{Schwinger}
\frac{i}{x+i\epsilon} \to \int_0^\infty d a e^{iax}, \qquad
 2\pi \delta(x) \to \int_{-\infty}^\infty d a e^{iax} .
\end{equation}
We now integrate in the following order: First over the common Schwinger parameter,  
then over the parameter corresponding to the delta function, and then over
the momentum $l_1$ in $D$ dimensions. For the integral we get
\begin{multline}
I  =
\frac{ -i\pi^\frac{D}{2}}{2 (2\pi)^{D}} \Gamma\left(2 - \frac{D}{2}\right)
\cdot\\
\int_0^\infty d \alpha_2  d \alpha_3 d \alpha_4 \; [
s \; \alpha_3(\alpha_2 + \alpha_3+ \alpha_4 -1) - u \alpha_2 \alpha_4 
-i\epsilon]^{\frac{D}{2}-2}\;.
\end{multline}
with the Mandelstam variables $s = 2p_1 p_2$ and $u= 2 p_2 p_3$.
We then substitute $\alpha_3 \to \alpha_3 \alpha_4$, we integrate over
$\alpha_2$, followed by $\alpha_4$, and finally expand about $D= 4$ dimensions. Eventually, we integrate over $\alpha_3$ and get
for the integral
\begin{equation} \label{res1}
I=
\frac{-i}{32 \pi^2}
\frac{1}{6} 
\left[
	\frac{s}{s+u} + 
	\frac{su}{(s+u)^2} \log\left(\frac{s}{-u}\right) 
\right] + {\cal O}(\epsilon) .
\end{equation}
For the color-ordered amplitude we have to consider all cyclic permutations
and we see that we get two equal contributions~\eqref{res1} and
two contributions with the exchange $s \leftrightarrow u$. In the sum the brackets~\eqref{res1} cancel exactly and we get the known result
\begin{equation} \label{result}
A_{4}^{\text{NLO}} (1^+, 2^+, 3^+, 4^+) =
\frac{-i}{16 \pi^2}
\frac{1}{6} 
\frac{ [1 2] [3 4] }{ \langle 1 2 \rangle \langle 3 4 \rangle}
+ {\cal O}(\epsilon). 
\end{equation}

\subsection{Sketch of the NNLO four-gluon amplitude}

Here we want to briefly sketch, how we may proceed to calculate the all-plus four-gluon amplitude at NNLO with the help of \eqref{ampfor}, that is, with two additional pairs
of forward particles. With two additional forward pairs we inevitably
encounter two gluons with negative helicity and the supersymmetry
relations mentioned in the last section do not hold. 
Therefore we construct
the amplitude directly from the three-gluon primitive amplitude~\eqref{primitive}.

Adding two pairs in the forward limit we have different contributions depending on the kind where the paired momenta appear, for instance,\\
\begin{tikzpicture}[scale=0.7]
\draw (0,0) -- (1,0);
\draw (1,0) -- (1,1);
\filldraw[black] (0,0) node[anchor=east] {$1^+$};
\filldraw[black] (1,1) node[anchor=south] {$-l_1^-$};
\filldraw[black] (1,0) circle (2pt);
\draw (1,0) -- (2,0);
\draw (2,0) -- (2,1);
\filldraw[black] (2,1) node[anchor=south] {$2^+$};
\filldraw[black] (2,0) circle (2pt);
\draw (2,0) -- (3,0);
\draw (3,0) -- (3,1);
\filldraw[black] (3,1) node[anchor=south] {$l_1^+$};
\filldraw[black] (3,0) circle (2pt);
\draw (3,0) -- (4,0);
\draw (4,0) -- (4,1);
\filldraw[black] (4,1) node[anchor=south] {$3^+$};
\filldraw[black] (4,0) circle (2pt);
\draw (4,0) -- (5,0);
\draw (5,0) -- (5,1);
\filldraw[black] (5,1) node[anchor=south] {$-l_2^-$};
\filldraw[black] (5,0) circle (2pt);
\draw (5,0) -- (6,0);
\draw (6,0) -- (6,1);
\filldraw[black] (6,1) node[anchor=south] {$4^+$};
\filldraw[black] (6,0) circle (2pt);
\draw (6,0) -- (7,0);
\filldraw[black] (7,0) node[anchor=west] {$l_2^+$};
\end{tikzpicture}
\hfill\eqnum\label{A422}\\
and contributions, where the vacuum pair momenta do not
appear separated, for instance,\\
\begin{tikzpicture}[scale=0.7]
\draw (0,0) -- (1,0);
\draw (1,0) -- (1,1);
\filldraw[black] (0,0) node[anchor=east] {$1^+$};
\filldraw[black] (1,1) node[anchor=south] {$-l_1^-$};
\filldraw[black] (1,0) circle (2pt);
\draw (1,0) -- (2,0);
\draw (2,0) -- (2,1);
\filldraw[black] (2,1) node[anchor=south] {$2^+$};
\filldraw[black] (2,0) circle (2pt);
\draw (2,0) -- (3,0);
\draw (3,0) -- (3,1);
\filldraw[black] (3,1) node[anchor=south] {$-l_2^-$};
\filldraw[black] (3,0) circle (2pt);
\draw (3,0) -- (4,0);
\draw (4,0) -- (4,1);
\filldraw[black] (4,1) node[anchor=south] {$3^+$};
\filldraw[black] (4,0) circle (2pt);
\draw (4,0) -- (5,0);
\draw (5,0) -- (5,1);
\filldraw[black] (5,1) node[anchor=south] {$l_1^+$};
\filldraw[black] (5,0) circle (2pt);
\draw (5,0) -- (6,0);
\draw (6,0) -- (6,1);
\filldraw[black] (6,1) node[anchor=south] {$4^+$};
\filldraw[black] (6,0) circle (2pt);
\draw (6,0) -- (7,0);
\filldraw[black] (7,0) node[anchor=west] {$l_2^+$};
\end{tikzpicture}
\hfill\eqnum\label{A423}\\
The latter class would, in terms of Feynman diagrams correspond
to non-planar contributions, which we can easily see connecting the
lines of opposite momenta.

For the unsplit subamplitudes~\eqref{ampfor} we get 
maximally helicity violating amplitudes which are given by the Parke-Taylor formula~\eqref{Parke-Taylor}, \cite{Parke:1986gb}.
Following~\eqref{ampfor} we have for the NNLO 
calculation to consider two additional pairs and up to two splittings of the subamplitudes, for instance a subamplitude with one splitting - originating from the subamplitude~\eqref{A423} - reads\\
\begin{tikzpicture}[scale=0.7]
\draw (0,0) -- (1,0);
\draw (1,0) -- (1,1);
\filldraw[black] (0,0) node[anchor=east] {$1^+$};
\filldraw[black] (1,1) node[anchor=south] {$-l_1^-$};
\filldraw[black] (1,0) circle (2pt);
\draw (1,0) -- (2,0);
\draw (2,0) -- (2,1);
\filldraw[black] (2,1) node[anchor=south] {$2^+$};
\filldraw[black] (2,0) circle (2pt);
\draw (2,0) -- (3,0);
\filldraw[black] (3,0) node[anchor=west] {$-l_3^-$};
\filldraw[black] (6,0) node[anchor=east] {$l_3^+$};
\draw (6,0) -- (7,0);
\draw (7,0) -- (7,1);
\filldraw[black] (7,1) node[anchor=south] {$-l_2$};
\filldraw[black] (7,0) circle (2pt);
\draw (7,0) -- (8,0);
\draw (8,0) -- (8,1);
\filldraw[black] (8,1) node[anchor=south] {$3^+$};
\filldraw[black] (8,0) circle (2pt);
\draw (8,0) -- (9,0);
\draw (9,0) -- (9,1);
\filldraw[black] (9,1) node[anchor=south] {$l_1^+$};
\filldraw[black] (9,0) circle (2pt);
\draw (9,0) -- (10,0);
\draw (10,0) -- (10,1);
\filldraw[black] (10,1) node[anchor=south] {$4^+$};
\filldraw[black] (10,0) circle (2pt);
\draw (10,0) -- (11,0);
\filldraw[black] (11,0) node[anchor=west] {$l_2^+$};
\end{tikzpicture}\\
\phantom{space}
\hfill\eqnum\label{A42s}\\
However, this partial amplitudes have to vanish because we do 
only have three external particles with minus helicities available and have to distribute them to
two partial amplitudes. Since all $n$-gluon tree amplitudes with less than
two minus helicities vanish, one partial amplitude of the product has to vanish.
Similar, for two splittings we have in total three partial amplitudes
and four pairs in the forward limit, that is, we have in total four 
 gluons with minus helicity available - however we would need
at least six minus helicities. 

Altogether we see that the NNLO four-gluon all-plus helicity
gluon amplitude is given by 
\begin{multline} \label{A4dec}
A_4^{\text NNLO}(1^+, 2^+, 3^+, 4^+) =\\
\int\limits_{l_1, l_2}\!\!\!
A_{4+4}^{\text{2 pairs, 0 splits }}\!\!(1^+, 2^+, 3^+, 4^+, -l_1^{-h_1}\!, l_1^{+h_1}\!, -l_2^{-h_1}\!, l_2^{h_1}\!) .
\end{multline}
The subamplitudes of~\eqref{A4dec} are all maximally helicity violating amplitudes~\eqref{Parke-Taylor}. 
We have to consider all permutations
of the positions of the two additional pairs. This gives contributions with different denominators in~\eqref{Parke-Taylor}. Moreover, the integration has to be done in $D$ dimensions. One possibility would be to use the 
four-dimensional helicity scheme as applied in the last section, where the external momenta are kept four dimensional and only the forward pairs are kept in $D=4-2\epsilon$ dimensions.
It is left for future work to do this calculation and show that the known result can be reproduced at NNLO.

\section{Conclusions}
The main point of this paper is the statement that on-shell amplitudes 
at any perturbation order can be represented by tree amplitudes with additional pairs of particles in the forward limit. 
 These contributions are indistinguishable from the leading order amplitudes and give higher order contributions. Over the phase-space of the momenta of the particles in the forward limit has to be integrated. Additional pairs of particles in the forward limit appear by the cut of propagators. It has been shown by adding systematically vacuum pairs in the forward limit that on-shell amplitudes are generated order by order in terms of trivalent tree subamplitudes. 

Applying the BCFW recursion relations we can decompose all the trivalent subamplitudes in terms of primitive $A_3$ amplitudes. These primitive amplitudes in turn are fixed by little-group scaling and the gauge group. 

In an explicit example we have shown the calculation of 
the four-gluon all-plus helicity amplitude at next-to-leading order. Starting with a primitive~$A_3$ amplitude we have first glued together the higher-point tree amplitudes, taking additional pairs of particles in the forward limit into account. Eventually we have performed the phase space integrations and reproduced the known result. 

 We have briefly sketched how at NNLO the all-plus four-gluon amplitud can be constructed. The explicit calculation at this order is left for future work. It would be very interesting to see this computation in detail, especially 
 with respect to the phase-space integrals of the gluons in the forward limit. 

With the proposed method we get the scattering amplitudes at any order based on first principles, that is, Lorentz invariance, locality, unitarity and the underlying gauge groups. We are not facing any gauge redundancies, virtual particles, ghosts, or polarization ‘‘tensors''. 
All amplitudes are eventually built by on-shell tree amplitudes. 

Amplitudes in the forward limit yield in general singular expressions. Therefore the calculation has to be regularized - here we have used dimensional regularization. Let us mention that the renormalization program is not changed at all compared to the conventional approach to calculate scattering amplitudes via Feynman diagrams. 

We expect that massive particles do not 
cause principal concerns. However, obviously the
calculation will be more complicated. This can be seen directly from the helicity formalism, see also appendix~\ref{app:spinor}:
The spinors follow from the two-by-two matrix defined as $p = p_\mu (\sigma^{\mu})$, which for massive particles is no longer of rank one, but will have full rank. The decomposition in terms of one pair of spinors therefore has to be replaced by two pairs of spinor vectors. For instance in~\cite{Arkani-Hamed:2017jhn} it has been shown how the helicity formalism has to be extended for the massive case. It would be interesting to see explicit calculations of amplitudes based on the here proposed methods with massive particles. 

Let us mention eventually that we find it quite striking that the perturbative expansion of scattering amplitudes can be deduced from first principles. 

\acknowledgments
The project was supported in part by the project Fondecyt with No.~1200641.

\appendix
\section{Spinor helicity conventions}
\label{app:spinor}

We follow the conventions of the Weyl formalism as outlined in~\cite{Elvang:2013cua}. 
We write the contravariant four-vector $(p^\mu) = (E, \tvec{p})^\trans$, 
the metric is $(\eta_{\mu \nu}) = \diag(-1,+1,+1,+1)$, and we arrange the unit matrix and the three Pauli matrices into one vector
$(\sigma^\mu) = (\unitmatrix_2, \tvec{\sigma})$ 
Then the following $2 \times 2$ matrix is formed:
\begin{equation}
p = p_\mu (\sigma^\mu) = 
\begin{pmatrix} -E + p_3 & p_1 - i p_2\\ p_1 + i p_2 & - E - p_3 \end{pmatrix}.
\end{equation}
For massless particles we have $\det(p) = p^\mu p_\mu = m^2 = 0$, that is,
$p$ has rank~1 and can be written as a product of two-component vectors, 
called spinors,
\begin{equation} \label{defspinor}
p_{a \dot{b}} 
= - | p ]_a \langle p |_{\dot{b}} 
\end{equation}
with indices $a, \dot{a}, b, \dot{b} \in \{1,2\}$. The minus sign is convention.
The massive case corresponds to a matrix~$p$ with full rank and has for instance 
been discussed in~\cite{Arkani-Hamed:2017jhn}.
We note that the dotted and undotted indices are simply different indices. 
In order to form Lorentz-invariant scalar products, antisymmetric 
$\epsilon$ symbols are introduced,
\begin{equation}
\epsilon^{ab} = \epsilon^{\dot{a} \dot{b}} = -\epsilon_{ab} = -\epsilon_{\dot{a} \dot{b}} = 
\begin{pmatrix} \phantom{+}0 & 1\;\;\\ -1 & 0\;\; \end{pmatrix}.
\end{equation}
The $\epsilon$ symbols obey
 \begin{equation}
 \epsilon_{ab} \epsilon^{bc} = \delta_a^c, \qquad
 \epsilon_{\dot{a} \dot{b}} \epsilon^{\dot{b} \dot{c}} = \delta_{\dot{a}}^{\dot{c}}\;.
 \end{equation}
Defining the spinors with upper indices
\begin{equation}
[p |^a \equiv \epsilon^{ab} |p ]_b , \qquad
|p \rangle^{\dot{a}} \equiv \epsilon^{\dot{a} \dot{b}} \langle p |_{\dot{b}}\;.
\end{equation}
 it can be shown that the following scalar products are Lorentz-invariant,
 \begin{equation}
 \begin{split}
 &\langle p q \rangle \equiv \langle p|_{\dot{a}}   \cdot | q \rangle^{\dot{a}} = 
 \epsilon^{\dot{a} \dot{b}}  \langle p|_{\dot{a}}   \cdot \langle q|_{\dot{b}}, 
 \\
& [p q ] \equiv [p |^a \cdot | q ]_a =
 \epsilon^{ab} |p]_b \cdot | q ]_a\;.
 \end{split}
 \end{equation}
We set 
 \begin{equation}
  p_{a \dot{b}}  \epsilon^{ab} \epsilon^{\dot{b} \dot{a}} =
  - | p ]_a \langle p |_{\dot{b}} \epsilon^{ab} \epsilon^{\dot{b} \dot{a}}  =
  - | p \rangle^{\dot{a}} [p|^b \equiv p^{\dot{a} b}
  \end{equation}
  and get
  \begin{equation}
   p_{a \dot{a}} q^{\dot{a} a} = -\langle pq \rangle [ pq ] = (p+q)^2 = 2 pq\;.
 \end{equation}
 Usually, the following scalar products are also defined,
\begin{equation}
\begin{split}
&\langle p | k | q] \equiv \langle p |_{\dot{a}} k^{\dot{a}b} |q]_b = -\langle pk \rangle [kq],
\\
&[p| k | q \rangle \equiv [p|^a k_{a \dot{b}} |q \rangle^{\dot{b}} = -[pk] \langle kq \rangle.
\end{split}
\end{equation}

Let us also mention some equations useful in the explicit calculation - see~\cite{Elvang:2013cua} for more details:
\begin{equation}
\langle ab \rangle  = - \langle ba \rangle, \qquad
[ab] = - [ba]
\end{equation}
and the Shouten identities
\begin{equation}
\begin{split}
&\langle ab \rangle \langle cd \rangle + 
\langle ac \rangle \langle db \rangle + 
\langle ad \rangle \langle bc \rangle = 0, 
\\
&[ab] [cd] +
[ac][db] +
[ad][bc] =0.
\end{split}
\end{equation}
For real momenta we have $[p|^a = (|p\rangle^{\dot{a}})^*$ and
$\langle p|_{\dot{a}}=( |p]_a)^*$ justifying the dot notation of the indices. 
Also useful is the context
\begin{equation}
|-p \rangle = -| p \rangle, \qquad
|-p ] = + |p].
\end{equation}


\bibliography{alltree}

\end{document}